\documentstyle[aps,floats,epsfig,axodraw]{revtex}
\input epsf
\def\beq{\begin{equation}}
\def\eeq{\end{equation}}
\def\bea{\begin{eqnarray}}
\def\eea{\end{eqnarray}}
\def\bq{\begin{quote}}
\def\eq{\end{quote}}
\def \lsim{\mathrel{\vcenter
     {\hbox{$<$}\nointerlineskip\hbox{$\sim$}}}}

\def\gappeq{\mathrel{\rlap {\raise.5ex\hbox{$>$}}
{\lower.5ex\hbox{$\sim$}}}}
\def\lappeq{\mathrel{\rlap{\raise.5ex\hbox{$<$}}
{\lower.5ex\hbox{$\sim$}}}}

\def\snuvi{\langle\tilde{\nu}_i\rangle}

\def\Rp{R_p}
\def\Rpv{R_p \! \! \! \! \! \! /~~}
\def\mnu{[m_{\nu}]_{ij}}

\def\vd{\vec{\delta}}

\newcommand{\sun}{\Delta m^2_{\rm solar}}
\newcommand{\atm}{\Delta m^2_{\rm atm}}

\def\bea{\begin{eqnarray}}   
\def\eea{\end{eqnarray}}

\begin{document}
\vspace*{-1in}
\renewcommand{\thefootnote}{\fnsymbol{footnote}}
\begin{flushright}
OUTP 01-49P.\\
LPT Orsay/01-81\\
CI-UAN 01-02T\\
\texttt{hep-ph/yymmddd} 
\end{flushright}
\vskip 5pt
\begin{center}
{\Large {\bf  Neutrino masses and mixings in the MSSM with soft 
 bilinear  $R_p$ violation
}}
\vskip 25pt
{\bf Asmaa Abada $^{1,}$\footnote{E-mail address:
abada@lyre.th.u-psud.fr}, Sacha Davidson $^{2,}$\footnote{E-mail address:
davidson@thphys.ox.ac.uk} and Marta Losada $^{3,}$}\footnote{E-mail address:
malosada@venus.uanarino.edu.co} 
 
\vskip 10pt  
$^1${\it Laboratoire de Physique Th\'eorique, 
Universit\'e de Paris XI, B\^atiment 210, 91405 Orsay Cedex,
France} \\
$^2${\it Theoretical Physics,  Oxford
University, 1 Keble Road, Oxford, OX1 3NP, United Kingdom}\\
$^3${\it Centro de Investigaciones, 
Universidad Antonio Nari\~{n}o, Cll. 59 No. 37-71, Santa Fe de Bogot\'{a},
Colombia}
\vskip 20pt
{\bf Abstract}
\end{center}

\begin{quotation}
  {\noindent\small 

\vskip 10pt
\noindent
We analyse a simple RPV extension of the MSSM, with bilinear
 R-parity  violation in the soft terms and vevs,
but not between the terms in the superpotential. The model gives two
massive neutrinos, and can fit all constraints from neutrino data.
We show analytically how to compute the lepton number violating
Lagrangian parameters from neutrino masses and mixing angles. Conversely,
we numerically vary the bilinear couplings as input parameters to
determine the allowed ranges that are consistent with neutrino
data. We briefly comment on the implications of our
bounds for low energy  LFV processes.
\\

PACS number(s):~12.60.Jv, 14.60.Pq, 11.30.Fs\\

}

\end{quotation}
\noindent{Neutrino Physics,
Supersymmetric Standard Model, Solar and Atmospheric Neutrinos}

\vskip 20pt  

\setcounter{footnote}{0}
\renewcommand{\thefootnote}{\arabic{footnote}}

\newpage
\section{Introduction}

The observed solar \cite{SNO,sun} and atmospheric neutrino deficits
\cite{ska,superk}  can be
explained by flavour-non-diagonal neutrino
masses. 
  The atmospheric neutrino anomaly is consistent with  
$\atm \sim (2 - 5) \times 10^{-3} ~{\rm eV}^2$ (with $\sin^2
2\theta_{\rm{atm}}> 0.88$) \cite{superk}.
The vacuum oscillation interpretation of solar neutrino data requires
$\sun \sim 10^{-10} ~{\rm eV}^2$, while the matter enhanced MSW
solution prefers the range $\sun \sim (10^{-10} - 10^{-4})~ {\rm eV}^2$
\cite{SNO,sun}, see table \ref{fitss}.  In the standard three 
neutrino framework  \footnote{The explanation of the LSND \cite{lsnd} data  
 requires a third
neutrino mass squared difference. LSND has not been confirmed and will be
checked by the future MiniBooNe \cite{Mboone}.}, 
which offers two independent mass differences, 
there are various mass hierarchies which can explain both
solar and atmospheric results. These are summarised
in table \ref{spectra}. 
The oscillation explanation
 of the solar neutrino problem and the atmospheric
neutrino anomaly  requires two  mixing angles and
 two  hierarchical  neutrino mass squared
differences, namely $\Delta m^2_{{\mathrm{sun}}} \ll \Delta m^2_{{\mathrm{atm}}}$.

\begin{table}[hbt]
\begin{tabular}{|c|l|l|l|}
Experiment & $\Delta m^2$ (eV$^2$) & $\sin^2 2\theta$ & $\tan^2\theta$\\\hline
Atmospheric & $(2-5)\times 10^{-3}$& $0.88-1$ & --\\\hline
 MSW-{LMA}&$(2-70)\times 10^{-5}$ & $0.6\ -\ 1$& $(2-40)\times 10^{-1}$\\
 MSW-{SMA}&$(0.4-1)\times 10^{-5}$ & $10^{-3}\ -\ 10^{-2}$& $(1-30)\times 10^{-4}$\\
 MSW-LOW&$ 4\times 10^{-10}-2\times 10^{-7}$ & $0.7\ -\ 1$ & $(1-80)\times 10^{-1}$\\
 Vacuum&$(1-6) \times 10^{-10}$ & $0.5\ -\ 1$& $(1-90)\times 10^{-1}$\\
 Just-so&$(4-10)\times 10^{-12}$ & $0.5\ -\ 1$ & $(3-30)\times 10^{-1}$\\\hline
Chooz & $> 3 \times 10^{-3}$& $\sin\theta<0.22$ \protect\label{fitss}
\end{tabular}
\vskip 0.5cm
\caption{Allowed mass squared differences and mixing angles for MSW-LMA, 
MSW-SMA, MSW-LOW,
Vacuum and Just-so  stand for MSW large mixing angle, small mixing, 
and  low, and for vacuum and just-so 
oscillation solutions, 
respectively. See e.g. \protect\cite{BCC}. We take the most conservative
bounds. 
}
\end{table}

\vskip 1cm

\begin{table}[hbt]
\begin{tabular}{|c|l|l|}
Spectrum  &Solar& Atmospheric \\ \hline
Hierarchical  &$\Delta m_{12}^2$& $\Delta m_{13}^{2}$ \\ \hline
Degenerate & $\Delta m_{23}^2 \quad{\mathrm{or}}\quad \Delta m_{12}^{2}$ & 
$\Delta m_{13}^{2}$ \\ \hline
Pseudo-Dirac & $\Delta m_{23}^2 \quad{\mathrm{or}}\quad \Delta m_{12}^{2}$ & 
$\Delta m_{13}^{2}$ \protect\label{spectra}
\end{tabular}

\vskip 0.5cm
\caption{Different possible regimes and corresponding  mass squared
difference. }\label{spectra}
\end{table}

We assume that the three neutrino masses  $m_{i}$, $ i =1,2,3$,  
satisfy  $m_1<m_2<m_3$.
As we will see, in our model we only have two non-zero  neutrino masses 
thus our spectra can only be  hierarchical or pseudo-Dirac, with
$\Delta m_{\mathrm atm}^{2} = \Delta m_{13}^{2} $ and 
$\Delta m_{\mathrm solar}^{2} = \Delta m_{12}^{2}, \Delta m_{23}^{2} $ 
respectively, see 
table \ref{spectra}.
 The $3 \times 3$ rotation   matrix \cite{Maki:1962mu} (MNS) 
which rotates from neutrino flavour
($f$) to  mass ($m$) eigenstates can be parametrised
by three rotations:
$V_{fm} = R_{23}(\theta_{23})R_{13}(\theta_{13})R_{12}(\theta_{12})$.
We neglect the Majorana and Dirac phases, 
and write
\beq
V_{fm} = \left[ \begin{array}{ccc} 
c_{12} c_{13} & c_{13} s_{12} &s_{13} \\
-c_{23} s_{12} - c_{12} s_{13} s_{23} 
& c_{12} c_{23} - s_{12} s_{13} s_{23} & c_{13} s_{23} \\
s_{23} s_{12} - c_{12} c_{23} s_{13} & 
-c_{12} s_{23} - c_{23} s_{12} s_{13} & c_{13} c_{23}
\end{array} \right].
\label{mixing}
\eeq

Massive neutrinos can be accomodated in the
 R-parity
($\Rp$) violating Supersymmetric Standard Model \cite{Nilles,hall,GH,dreiner}, where
$\Rp$ is defined as $(-1)^{3B+L +2S}$,
and  $B,L$ and $S$ are respectively
the baryon number, lepton number and spin.
$\Rp$ is 
frequently imposed
on the Supersymmetric SM to
forbid renormalisable
baryon and lepton number violating
interactions. However, the
phenomenological bounds
on $B$ and/or $L$ violation \cite{dreiner}
can 
be satisfied by
imposing $B$ as a symmetry
and 
allowing the lepton number violating couplings
to be large enough to generate Majorana
neutrino masses.
We assume in this work that
the baryon number violating couplings
are absent, and
concentrate on the ``bilinear''
lepton number violating
interactions in the superpotential and soft terms.
 We neglect the  trilinear  R-parity
violating (RPV) couplings that can appear in the superpotential.

Neutrino masses
in RPV  theories have been studied analytically
for many years \cite{hall,everyone}, and recently
numerically with more care \cite{num,chunetal,hirschetal}.
The calculation of the neutrino masses to any given order 
can be performed  by calculating the relevant 
self energy diagrams in the mass insertion approximation
for the small  RPV mass parameters. Alternatively, one
can diagonalise the RPV mass matrices 
of particles which propagate in the  loops.
We use the mass insertion approximation, which allows us to 
identify {\it analytically} the contribution from the different parameters
and thus use neutrino data to constrain the RPV couplings directly.
 Our purpose in this paper is to consider a simple model in which 
lepton number violation appears as a misalignment
of the soft terms with respect to the superpotential. We then calculate 
the neutrino mass matrix  in the mass insertion approximation.

Constraints on  RPV parameters from neutrino masses and mixings 
are significant,  as typically they provide the
 stronger bounds  on
the couplings than other low energy lepton-flavor violating (LFV) processes.
In this paper we present two approaches in order to constrain the bilinear 
R-parity violating Lagrangian parameters.
The first one is an analytic approach in which we  obtain expressions for the
 two
bilinear RPV parameters appearing in the Lagrangian in terms of  neutrino masses and mixing angles.
The second approach  is a numerical one which follows that in ref \cite{AM1,AM2}. The bilinear parameters
are varied as inputs and we use neutrino data to constrain the allowed ranges of these parameters.
For that, since we only  have ranges 
for the physically observed inputs, we  constrain our parameter space (6
variables) 
  by using the  following  inputs from neutrino data: 
\begin{itemize}
\item $\Delta m^2_{\mathrm{atm}}$ and $\sin ^2 2\theta_{\mathrm{atm}}$,
\item  $\Delta m^2_{\mathrm{sun}}$ and $\sin ^2 2\theta_{\mathrm{sun}}$,
\item $\sin\theta_{{\mathrm{Chooz}}}$.
\end{itemize}
Imposing these constraints implies an indirect restriction on the effective 
mass 
$m_{\mathrm{eff}}$ that enters the  neutrinoless-double beta decay amplitude, i.e., 
$|m_{\mathrm{eff}}|\le\sum_{i}^{}m_{\nu_i}|V^2_{ei}|$. As we will show, we have
 checked in the different 
analyses we did of
that the  experimental bound  \cite{76Ge} on $m_{\mathrm{eff}}$ is always satisfied.\\
The paper is organized as follows, in section II we present the model. 
Section III presents analytic and numerical
results. 
Section IV discusses a model which generates neutrinos 
masses at tree-level and through loops considering 
only $R_p$-violating parameter contributions from the misalignment
 of the $\mu$ parameters and the vevs. In section V we  discuss the implications of our
results for low energy lepton flavour violating processes.
We then present a summary of the paper.

\section{The Bilinear RPV Model}

In 
a $\Rpv$ supersymmetric model, the
Higgs and the sleptons can mix
(they have the same gauge quantum numbers), 
so  the down-type
Higgs and sleptons can be assembled in a vector $L_J = (H_d, L_i)$ 
with $J:4..1 $. 
With this notation, 
the superpotential for the  supersymmetric
SM with   $R_p$ violation  can be written as

\beq
W= \mu^J {H}_u  L_J + \frac{1}{2}\lambda^{JK \ell} L_J L_K E^c_{\ell} + 
\lambda^{'Jpq} L_JQ_p D^c_q  + h_u^{pq} {H}_u Q_p U^c_q ~~.\label{S}
\eeq
The $R_p$ violating and
conserving coupling constants
have been assembled into vectors and matrices in $L_J$ space:
we call the usual $\mu$ parameter $\mu_4$,
and identify the usual $\epsilon_i =  \mu_i$, 
$h_e^{jk} = \lambda^{4jk}$, and
$h_d^{pq} = \lambda^{' 4pq}$.
Lower case roman indices $i,j,k$ and $p,q$  are lepton and
quark generation indices.
We also include possible $R_p$ violating couplings among
the soft SUSY breaking parameters, which can
be written as
\begin{eqnarray}
V_{soft}& = & \frac{\tilde{m}_u^2}{2} H_u^{\dagger} H_u + \frac{1}{2}
 L^{J \dagger} [\tilde{m}^2_L]_{JK} L^K  + B^J H_u L_J   \nonumber \\ 
& & + A_u^{ps} H_u Q_p U^c_s + 
     A^{'Jps} L_J Q_p D^c_s +
     \frac{1}{2}A^{JKl} L_J L_K E^c_l + h.c.~~~.  \label{soft}
\end{eqnarray}

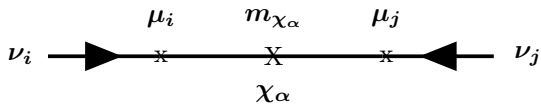
\begin{figure}[htb]
\unitlength1mm
\SetScale{2.8}
\begin{boldmath}
\begin{center}
\begin{picture}(60,30)(0,-10)
\ArrowLine(0,0)(15,0)
\Line(45,0)(15,0)
\ArrowLine(60,0)(45,0)
\Text(-2,0)[r]{$\nu_i$}
\Text(62,0)[l]{$\nu_j$}
\Text(30,-5)[c]{$\chi_{\alpha}$}
\Text(30,5)[c]{$m_{\chi_{\alpha}}$}
\Text(30,0)[c]{X}
\Text(15,0)[c]{x}
\Text(45,0)[c]{x}
\Text(15,5)[c]{$\mu_i$}
\Text(45,5)[c]{$\mu_j$}
\end{picture}
\end{center}
\end{boldmath}
\caption{Tree-level neutrino mass in the mass
insertion approximation.}
\protect\label{tree}
\end{figure}

Clearly field redefinitions of the $H_{d}, L_{i}$ fields correspond to basis changes 
in $L_{J}$ space and consequently the Lagrangian parameters
will be altered. Thus, whenever constraints are placed on the Lagrangian 
parameters the basis in which these are valid must be defined.
Alternatively, we can construct {\bf basis-independent} parametrisations of 
the couplings and constrain these. In this paper we take the second
approach using the basis independent parameters $\delta_{\mu}, \delta_{B}$, 
constructed in refs.\cite{DL1,DL2}, which are defined in table \ref{tinvar}.\\
It is well known that this model can generate  $\Delta L=2$ neutrino 
masses from tree-level and loop diagrams.
There will be a non-zero
neutrino  mass at tree-level 
if $\mu_I \neq (\sqrt{ \sum_J \mu_J^2}/\sqrt{\sum_J
v_J^2}) v_I$---that is if $(\mu_4,\mu_3,\mu_2,\mu_1)$ 
is not parallel with
$(v_4,v_3,v_2,v_1)$ \cite{Nilles}. 
In the $v_i = \snuvi = 0$ basis,
a neutrino $\nu_3$ acquires mass $m_3$ at tree
level via a ``seesaw'', with neutralinos
playing the role of the heavy Majorana fermion,
and the mass $\mu_i \nu_i \tilde{h}_u^o $
in the place of the ``Dirac'' mass, see figure \ref{tree}.

\begin{figure}[htb]
\unitlength1mm
\SetScale{2.8}
\begin{boldmath}
\begin{center}
\begin{picture}(60,50)(0,-10)
\Line(0,0)(15,0)
\Line(45,0)(15,0)
\Line(60,0)(45,0)
\DashCArc(30,0)(15,0,180){1}
\Text(8,0)[c]{$\bullet$}
\Text(8,3)[c]{$I$}
\Text(20,11)[c]{$\bullet$}
\Text(16,14)[c]{$IV$}
\Text(52,0)[c]{$\bullet$}
\Text(52,3)[c]{$VIII$}
\Text(40,11)[c]{$\bullet$}
\Text(43,14)[c]{$VI$}
\Text(-2,0)[r]{$\nu_i$}
\Text(62,0)[l]{$\nu_j$}
\Text(30,20)[c]{$h,H,-A$}
\Text(30,0)[c]{x}
\Text(30,-6)[c]{$\chi$}
\end{picture}
\end{center}
\end{boldmath}
\caption{
The Grossman-Haber loop. The blobs
indicate possible positions for $\Rpv$  mass insertions.
The misalignment between $\vec{\mu}$ and
$\vec{v}$ allows a mass insertion on
the lepton/higgsino lines (at points
I,  or VIII). The
soft $\Rpv$ masses appear as mass insertions
at positions VI and IV on
the scalar line. } 
\label{fGH}
\end{figure}
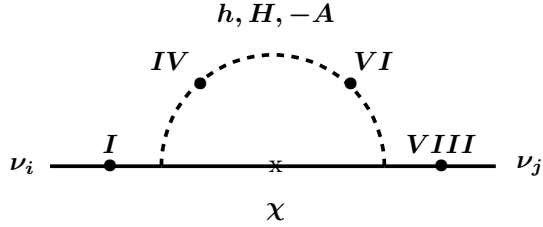

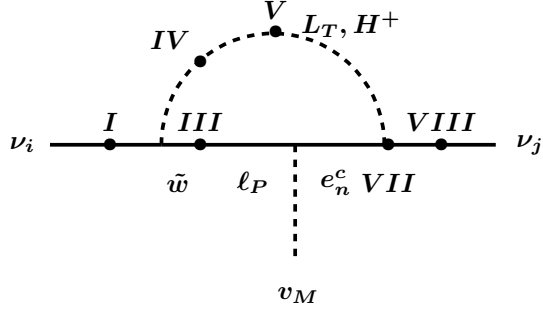
\begin{figure}[htb]
\unitlength1mm
\SetScale{2.8}
\begin{boldmath}
\begin{center}
\begin{picture}(60,60)(0,-20)
\Text(8,0)[c]{$\bullet$}
\Text(8,3)[c]{$I$}
\Text(20,0)[c]{$\bullet$}
\Text(20,3)[c]{$III$}
\Text(45,0)[c]{$\bullet$}
\Text(45,-5)[c]{$VII$}
\Text(52,0)[c]{$\bullet$}
\Text(52,3)[c]{$VIII$}
\Text(20,11)[c]{$\bullet$}
\Text(16,14)[c]{$IV$}
\Text(30,15)[c]{$\bullet$}
\Text(30,18)[c]{$V$}
\Line(0,0)(15,0)
\Line(45,0)(15,0)
\Line(60,0)(45,0)
\DashCArc(30,0)(15,0,180){1}
\Text(-2,0)[r]{$\nu_i$}
\Text(62,0)[l]{$\nu_j$}
\Text(40,16)[c]{$L_T, H^+$}
\Text(17,-5)[c]{$\tilde{w}$}
\Text(27,-5)[c]{$\ell_P$}
\Text(38,-5)[c]{$e^c_n$}
\Text(33,-20)[c]{$v_M$}
\DashLine(33,0)(33,-15){1}
\end{picture}
\end{center}
\end{boldmath}
\caption{Schematic representation
of the charged loops
 with one gauge and  a Yukawa coupling. 
These diagrams occur
if gauginos mix with  charged leptons---that is if
$\delta_{\mu} \neq 0$. 
The blobs
indicate possible positions for $\Rpv$ interactions;
each diagram contains two blobs.  
 The $\Rpv$ interactions can be 
trilinears (at  VII) or mass insertions.
The misalignment between $\vec{\mu}$ and
$\vec{v}$ allows a mass insertion on
the lepton/higgsino lines (at points
I, III, or VIII), or on the
scalar line at point $V$. The
soft $\Rpv$ masses can appear as a mass insertion
at position IV. } 
\label{fglambda}
\end{figure}

In the mass insertion
approximation for $\Rpv$ masses
there are three types of loops
that can contribute masses to the two neutrinos
that are massless at tree level.
There are the well-known loops
involving trilinear $\Rpv$ 
couplings
$\lambda$ or $\lambda'$ at both vertices. 
Then there is the Grossman-Haber
diagram \cite{GH}, figure \ref{fGH}, with gauge
couplings at the vertices and  $\Rpv$ 
provided by mass insertions of the
soft masses which mix the Higgses and
the sleptons. Finally, there is figure \ref{fglambda}, which has
one gauge coupling and one trilinear/yukawa,
and at least one unit of lepton number violation
provided by a $\Rpv$ mass insertion. 
The last two types of loops have frequently been
neglected in analytic estimates of
neutrino masses. 
In a previous work \cite{DL1,DL2}, 
two of us showed that the analytic
estimates in models
with lepton number violating masses ($\equiv$  bilinear $\Rpv$)
suffer from two linked
problems: confusion about the interaction
eigenstate basis choice in the Lagrangian,
and an incomplete set of
one loop diagrams contributing
to neutrino masses. ``Basis independent''
formulae for all the diagrams can be
found in \cite{DL2}. 
For a discussion of the construction
of basis-independent invariants, see,
{\it e.g.} \cite{Nilles,DL1,DL2,sacha,Fer,DLR,GH3}.

R-parity violating models can be divided
into classes based on which of the
diagrams make significant contributions
to the neutrino mass matrix $\mnu$ \cite{DL1,DL2}:
in the first case, the loop diagrams with bilinear $\Rpv$
are much smaller than the trilinear
diagrams. In this case, $\mnu$
is due to the tree mass and the canonical
$\lambda$ and $ \lambda'$ diagrams,
and can have a large variety of
patterns depending on the MSSM and $\Rpv$ inputs.
This case has recently been carefully studied in
refs. \cite{AM1,AM2}.  In the second case, the
loops with bilinear $\Rpv$ due to soft masses make significant
contributions to $\mnu$, but the bilinears contributing
at tree level$(\delta_{\mu}^i)$ are negligible in
the loops. This is the case we study in this  section.
To isolate the effect of the soft $\Rpv$, we
assume that the trilinears are negligible. 
The third possibility is that loops with lepton number
violation from $\delta_{\mu}$ are important.
Such loops can contribute $\mnu \sim \delta_{\mu}^2 h_\tau^2 m_\tau^2/
(16 \pi m_{SUSY})$, whereas the tree contribution
is $m_{\nu}^{tree} \sim \delta_{\mu}^2 m_{SUSY}$.
This generates  a very  large hierarchy among
neutrino masses, so this case is unlikely to
be realised. We discuss this further in section \ref{sec3}.

\begin{table}
\begin{center}$
\setlength{\arraycolsep}{2em}
\begin{array}{||c|c||}
\hline
\hline
 & \\
\delta_{\mu}^i  \equiv \frac{ \vec{\mu} \cdot  \lambda^i \cdot  \vec{ v}}{|  \vec{\mu}| \sqrt{2} m_e^i} 
&
\frac{\mu^i}{|\mu|} \\
 & \\
\hline
 &  \\
\delta_{B}^i \equiv \frac{  \vec{B} \cdot  \lambda^i  
  \cdot  \vec{v}}{|  \vec{B}|  \sqrt{2} m_e^i} & \frac{B^i}{|B|}\\
 &  \\
\hline
 &  \\
\delta_{\lambda'}^{ipq} \equiv \frac{  \vec{\lambda}^{'pq}  \cdot 
\lambda^i  \cdot   \vec{v}}
{ \sqrt{2} m_e^i} & {\lambda}^{'ipq}\\
 &  \\
\hline
 &  \\
\delta_{\lambda}^{ijk} \equiv \frac{ \vec{v}  \cdot \lambda^{i}
\lambda^{k} \lambda^j  \cdot  \vec{v}}
{2 m_e^i m_e^j} &{\lambda}^{ijk} \\
 &  \\
\hline
\hline
\end{array}
$\end{center}
\caption{The basis-independent invariants used
to parametrise the bilinear $\Rpv$ relevant for
neutrino masses, together with
their value in the $\snuvi = 0$ basis. 
They are zero if $R_p$ is conserved. 
(Note
that these invariants have signs: for arbitrary vectors
 $ \vec{a}$ and   $ \vec{b}$,
  $ \vec{a} \cdot \lambda^{i}  \cdot \vec{b} = -
  \vec{b}  \cdot  \lambda^{i}  \cdot \vec{a} $).}
\label{tinvar}
\end{table}

We take in this section a model where
$\delta_{B}^i, \delta_{\mu}^j \neq 0$,
and where $\delta_{\lambda}, \delta_{\lambda'}$ 
are negligible. 
Such a model could arise if the $\Rpv$ is originated
in the soft terms. It is interesting
to consider because it shows the contribution
of $\delta_{B}^j$ to  $\mnu$, and allows
us to set bounds on the $\delta_B^j$. We
want to determine which values of
$(\delta_B^1,\delta_B^2,\delta_B^3)$
and $(\delta_{\mu}^1,\delta_{\mu}^2,\delta_{\mu}^3)$
reproduce a neutrino
mass matrix consistent with the data. 
We do this in two ways. First
we assume that
the neutrino masses and mixing angles are
all known, and analytically solve for
$\vec{\delta}_\mu$ and 
$\vec{\delta}_B$ as a function of the 
masses and mixing angles. This is
tractable because the $\vec{\delta}_\mu$, 
$\vec{\delta}_B$
model only has two non-zero neutrino masses,
so the algebra reduces to two
dimensions. 

Secondly, we start
with $\vec{\delta}_\mu$ and 
$\vec{\delta}_B$ as inputs, varied over ``sensible''
ranges, and numerically determine
whether the resulting
neutrino mass matrix is consistent
with observations using the inputs given above.

There are three relevant diagrams contributing to
$\mnu$ in this model: the tree
diagram of figure \ref{tree}, and the Grossman-Haber diagram
of figure \ref{fGH} with $\Rpv$ at points
IV and VI, and with  $\Rpv$ at points
I and VI. Exact formulae for these
diagrams can be found in \cite{DL2}.
Our simplifying assumptions  generate a neutrino mass matrix of the
form
\beq
\mnu = m_\mu^{(0)} \hat{\delta}_{\mu}^i \hat{\delta}_{\mu}^j
+ m_{\mu B}^{(0)} (\hat{\delta}_{\mu}^i \hat{\delta}_{B}^j +
\hat{\delta}_{B}^i \hat{\delta}_{\mu}^j)
+  m_{ B}^{(0)}\hat{\delta}_{B}^i \hat{\delta}_{B}^j
\eeq
where 
\bea
m_{\mu}^{(0)} & = & |\vd_{\mu}|^2 m_{SUSY} \\
m_{\mu B}^{(0)} & = & \frac{\alpha}{16 \pi} |\vd_{\mu}| |\vd_{B}| m_{SUSY} 
=\sqrt{\frac{\alpha}{16 \pi} m_{\mu}^{(0)} m_{B}^{(0)}} \label{mmuB} \\
m_{B}^{(0)} & = &\frac{\alpha}{16 \pi} |\vd_{B}|^2 m_{SUSY} 
\eea
and $\alpha=\frac{g^2}{16\pi}$,  $\hat{\delta}_{\mu}$ and 
$\hat{\delta}_{B}$ are unit vectors in $\{ L^i \}$ space:
\beq
\hat{\delta}_{\mu} = \frac{1}{\sqrt{\sum_i (\delta_{\mu}^i)^2}}
( \delta_{\mu}^e, \delta_{\mu}^{\mu}, \delta_{\mu}^{\tau}) ~~~.
\eeq 

We have set all the unknown sparticle masses
equal to $m_{SUSY}$ = 100 GeV, and neglected
the mixing angles among MSSM particles.
This is phenomenologically reasonable since superpartners
 have not been detected. Using the correct dependence
on the MSSM parameters would have two effects:
$m_{\mu B}^{(0)}$ would no longer
be related to $ m_\mu^{(0)}$ and $m_{B}^{(0)}$
as in eqn (\ref{mmuB}), and there
could be additional
(probably small) contributions to
$\mnu$ with more complex index
structure induced by spartner mass differences.

\section{Results}

\subsection{$\vec{\delta}_\mu$ and 
$\vec{\delta}_B$ as a function of 
masses and mixing angles}

\label{dmudB}

In this section, we wish to go from
the data to the Lagrangian parameters
$\vec{\delta}_{\mu}$ and  $\vec{\delta}_{B}$.
We assume that the three neutrino masses (one of
which is zero) and
three mixing angles are all known exactly; from
these six inputs we wish to analytically solve for
the six parameters $\delta_{\mu}^i$ and  $\delta_{B}^i$, $(i=1,3)$. 
The  data do not fully determine
$\vec{\delta}_{\mu}$ and  $\vec{\delta}_{B}$,
because these two vectors are not required to
be orthogonal.  An orthonormal basis would be
$\hat{\delta}_{\mu}$ and $\hat{\delta}_{B_\perp}$
where
\beq
\hat{\delta}_{B_\perp} = \frac{1}{\sqrt{\sum_i (\delta_{B_\perp}^i)^2}}
( \delta_{B_\perp}^e, \delta_{B_\perp}^{\mu}, \delta_{B_\perp}^{\tau}) ~~~,
\eeq 
and
\beq
\delta_{B_\perp}^i =\delta_{B}^i -(\delta_{B} \cdot \hat{\delta}_{\mu})
  \hat{\delta}_{\mu}^i  ~~~.
\label{Bperp}
\eeq
The parameters 
$\vec{\delta}_{\mu}$ and  $\vec{\delta}_{B}$
can be expressed in terms of the two
neutrino masses and the mixing angles, and
one additional parameter which could be taken
as the angle between
$\vec{\delta}_{\mu}$ and  $\vec{\delta}_{B}$ 
(which we will call $\rho$).

We
first express the neutrino mass matrix $\mnu$ in terms of the orthonormal
basis  ($\hat{\delta}_{\mu}$, 
 $\hat{\delta}_{B_\perp}$).
Since  $\mnu$
is a mass matrix in the two dimensional space spanned
by $\vec{\delta}_{\mu}$
and  $\vec{\delta}_{B}$, 
it has two non-zero
eigenvalues and the problem reduces to
two dimensional algebra. Secondly,
we diagonalise $\mnu$,
and compare the eigenvalues and
eigenvectors to
the observed masses and mixing angles. This
allows us to determine $\vec{\delta}_{\mu}$
and  $\vec{\delta}_{B}$
as a function of the masses and mixing angles,
and one free parameter. The free
parameter  arises because $\vec{\delta}_{\mu}$
and  $\vec{\delta}_{B}$ are not orthogonal, so what
fraction of the heavier neutrino mass is
due to  $\vec{\delta}_{B}$ is a free parameter.

In the ($\hat{\delta}_{\mu}$, $\hat{\delta}_{B_\perp}$)
basis, $\mnu$ can be written
\bea
\mnu &=& (m_\mu^{(0)}  +  2 m_{\mu B}^{(0)}\cos \rho  +  m_{ B}^{(0)}
\cos^2 \rho)
\hat{\delta}_{\mu}^i \hat{\delta}_{\mu}^j  \nonumber \\ &&
+ (m_{\mu B}^{(0)} \sin \rho + m_B^{(0)} \sin \rho \cos \rho) 
(\hat{\delta}_{\mu}^i \hat{\delta}_{B_\perp}^j +
\hat{\delta}_{B_\perp}^i \hat{\delta}_{\mu}^j)
+  m_{ B}^{(0)}\sin^2 \rho 
\hat{\delta}_{B_\perp}^i \hat{\delta}_{B_\perp}^j \\
 & \equiv & m_\mu
\hat{\delta}_{\mu}^i \hat{\delta}_{\mu}^j  
+ m_{\mu B} (\hat{\delta}_{\mu}^i \hat{\delta}_{B_\perp}^j +
\hat{\delta}_{B_\perp}^i \hat{\delta}_{\mu}^j)
+  m_{ B} \hat{\delta}_{B_\perp}^i \hat{\delta}_{B_\perp}^j
\label{muB}
\eea
where $ \rho$ is the angle between $ \hat{\delta}_{\mu}$
and $ \hat{\delta}_{B}$ : $\cos \rho =   \hat{\delta}_{\mu} \cdot 
 \hat{\delta}_{B}$. The matrix 
$\mnu$ has two non-zero eigenvalues,
so the neutrino mass
spectrum is either
hierarchical with $\Delta m^2_{atm} = m_3^2$
and  $\Delta m^2_{sol} = m_2^2$, or
pseudo-Dirac, with $\Delta m^2_{atm} = m_3^2, m_2^2 $
and  $\Delta m^2_{sol} =  m_3^2-m_2^2$. 
The  eigenvectors are the last
two columns of $V_{fm}$ in eq. (\ref{mixing}), and lie in the 2-d
sub-space spanned by  
$(\delta_B^1,\delta_B^2,\delta_B^3)$
and $(\delta_{\mu}^1,\delta_{\mu}^2,\delta_{\mu}^3)$.

 We can now
determine $(\delta_B^1,\delta_B^2,\delta_B^3)$
and $(\delta_{\mu}^1,\delta_{\mu}^2,\delta_{\mu}^3)$
 as a function of the neutrino
masses and mixing angles, and  the angle $\gamma$
between  $(\delta_{\mu}^1,\delta_{\mu}^2,\delta_{\mu}^3)$
and $(V_{13},V_{23},V_{33})$.  
We solve for the angle $\rho$ (between
 $\vec{\delta}_B$  and $\vec{\delta}_\mu$)
as a function of $\gamma$ in equation (\ref{rho}).

In the mass eigenstate basis spanned by $V_{f2}$ and $V_{f3}$
($f:1..3)$, the neutrino mass matrix is
\beq
\left[\begin{array}{cc} 
m_2 & 0 \\
 0 & m_{ 3}\label{diago}
\end{array}
\right].
\eeq
Suppose that  $(\delta_{\mu}^1,\delta_{\mu}^2,\delta_{\mu}^3)$
is rotated by an  angle $\gamma$ with
respect to  $(V_{13},V_{23},V_{33})$, so that  the mass matrix (\ref{diago}) is
\beq
\left[\begin{array}{cc} 
\cos^2 \gamma m_2 + \sin^2 \gamma m_3 &  
\cos \gamma  \sin \gamma (m_2 - m_3)  \\
\cos \gamma  \sin \gamma (m_2 - m_3)  & 
\cos^2 \gamma m_3 + \sin^2 \gamma m_2
\end{array}
\right]
=
\left[\begin{array}{cc} 
m_B &  m_{\mu B} \\
 m_{\mu B} & m_{\mu}
\end{array}
\right],
\label{**}
\eeq
in the basis of
$(\delta_{\mu}^1,\delta_{\mu}^2,\delta_{\mu}^3)$
and $(\delta_{B_\perp}^1,\delta_{B_\perp}^2,\delta_{B_\perp}^3)$.
By comparing eqn (\ref{**}) to eqn (\ref{muB}), we find
\beq
m_B^{(0)} \sin^2 \rho \equiv  \frac{\alpha}{16 \pi} 
  |\delta_{B}|^2 m_{SUSY} \sin^2 \rho =
(\cos^2 \gamma m_2 + \sin^2 \gamma m_3)
\label{mb0}
\eeq

\beq
m_{\mu}^{(0)}\left( 1 - \frac{\alpha}{16 \pi} \right)
\equiv  |\delta_{\mu}|^2 m_{SUSY}\left( 1 - \frac{\alpha}{16 \pi} \right) = 
\frac{m_2 m_3}{\cos^2 \gamma m_2 + \sin^2 \gamma m_3}
\eeq

\beq 
\cot \rho = \frac{ \cos \gamma \sin \gamma (m_2 - m_3) }
{ \cos^2 \gamma m_2 + \sin^2 \gamma m_3 } - \sqrt{  \frac{\alpha}{16 \pi} 
\frac{m_2 m_3}{(\cos^2 \gamma m_2 + \sin^2 \gamma m_3)^2}}
\label{rho}
\eeq

\beq
\frac{1}{|\delta_{\mu}|}\delta_{\mu}^f =
\cos \gamma V_{f3} + \sin \gamma V_{f2}
\label{delm}
\eeq
\beq
\frac{1}{|\delta_{B_{\perp}}|}\delta_{B_{\perp}}^f
=-\sin \gamma V_{f3} +\cos \gamma V_{f2}.
\label{delBperp}
\eeq
We can now determine the values
of $(\delta_{\mu}^1,\delta_{\mu}^2,\delta_{\mu}^3)$
and $(\delta_{B}^1,\delta_{B}^2,\delta_{B}^3)$
corresponding to  atmospheric and different
solar solutions.

\begin{figure}[ht]
\begin{center}
\hspace{1cm}\epsfxsize=14cm\epsfbox{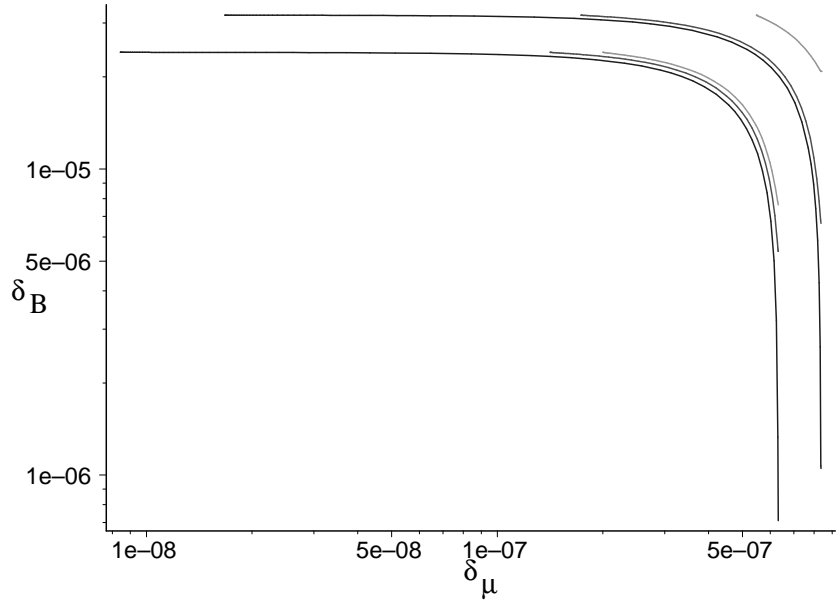}
\caption{Allowed region in parameter space for various solar
oscillation solutions combined with the atmospheric and Chooz constraints.
 The allowed region is between the lines; the
longest pair of lines correspond to the vacuum solar solution, and  
the shortest
[middle length] lines to the LMA [SMA] MSW solution.}
\protect\label{fLMA}
\end{center}\end{figure}

We first compute values of $|\delta_{\mu}|$
and $|\delta_{B}|$ that give solar and atmospheric
neutrino masses. \\To get degenerate masses 
$m_2 \sim m_3 \sim \sqrt{\Delta m^2_{atm}}$,
the angle $\rho$ between $\vec{\delta}_B$
and $\vec{\delta}_{\mu}$ must be of
order $ \pm \pi/2$. So $\vec{\delta}_B$
must be approximately orthogonal  to  $\vec{\delta}_{\mu}$,
and $\frac{\alpha}{16 \pi} |{\delta}_B|^2 \simeq 
 |{\delta}_\mu|^2 \simeq \sqrt{\Delta m^2_{atm}}/m_{SUSY}$.
This seems unnatural and fine-tuned, so
we do not pursue this possibility. \\
For the case of hierarchical neutrino masses,
we plot in figure \ref{fLMA} the allowed values of 
$|\delta_{\mu}|$ and $|\delta_B|$ corresponding to combined 
constraints from 
Chooz and atmospheric (SuperK) results
and three of the solar solutions. 
The allowed parameter space corresponds to the
space contained between the lines in  figure  \ref{fLMA}.
The shortest pair of lines corresponds to
the LMA MSW solution, the middle lines 
to the SMA solution and the longest pair of
lines to the Vacuum solution. 
We obtain these  two lines by varying the angle $\gamma$
for the maximum and minimum allowed values
of the masses $m_2$ and $m_3$, as given in
table \ref{fitss}.
In this plot (fig.\ref{fLMA}), the left [bottom] end
of the allowed parameter space corresponds
to $m_3$ induced by  ${\delta}_B$ [${\delta}_{\mu}$].

It was pointed out in \cite{hirschetal} that
their MSUGRA bilinear model was effectively controlled
by two vectors,  $\vec{\delta}_\mu$
and the misalignment between $\mu$
and the trilinears ($\equiv \delta_\lambda$). The parametrisation
of equations (\ref{mb0}) to (\ref{delBperp})
should  therefore  approximately describe 
the results of \cite{hirschetal},  replacing
$\vec{\delta}_B$ by $\vec{\delta}_\lambda$. Their heavy
neutrino mass $m_3$ is induced by the tree
diagram of figure \ref{tree}, so the angle 
$\gamma$ (between the eigenvector corresponding to $m_3$
and $\vec{\delta}_\mu$)   is small.  It is easy
to see from equation (\ref{delm}) that the
atmospheric mixing angle is
therefore  controlled by the relative
size of $\delta_\mu^\mu$ and  $\delta_\mu^\tau$,
as was indeed found by \cite{hirschetal}. 
We see from equation (\ref{delBperp})
that the solar mixing angles are determined
from the relative size of the 
elements of $\vec{\delta}_{B_\perp}$,
rather than  from the elements of $\vec{\delta}_B$.
This means that the solar angle
will sometimes show a correlation with 
the elements of  $\vec{\delta}_B$,
specifically  if  $\vec{\delta}_B$ is
approximately perpendicular to
$\vec{\delta}_\mu$. This agrees with
the results of \cite{hirschetal}.

The Chooz results\footnote{The original two-neutrino analysis of the Chooz
 data combined with 
the SuperK best fit atmospheric value gives
$|U_{e3}|^2\lsim 4.~10^{-2}$. In the framework of three-neutrino  mixing, the analysis 
of Chooz combined with the SuperK atmospheric data gives $|U_{e3}|^2\lsim 2.~10^{-2}$,
see ref. \cite{Bilenky-SM} and references therein.}
require $V_{e3}=  \cos \gamma  \hat{\delta}_\mu^e -
\sin \gamma \hat{\delta}_{B_\perp}^e \lappeq .1$.
For large angle solar solutions (LMA), $V_{e2} \sim 1/\sqrt{2}$,
so $\hat{\delta}_{\mu}^e \simeq \frac{1}{\sqrt{2}}\sin \gamma$,
and 
 $\hat{\delta}_{B_{\perp}}^e \simeq \frac{1}{\sqrt{2}}\cos \gamma$,
and $V_{e3} $ is small due to a cancellation
between   $\hat{\delta}_{\mu}^e$ and  $\hat{\delta}_{B_{\perp}}^e$.
For small angle solar solutions,
both  $\hat{\delta}_{\mu}^e$ and  $\hat{\delta}_{B_{\perp}}^e$
may be small. This scenario has
cosmological attractions,
because  it would allow a baryon asymmetry
produced before the electroweak
phase transition to survive. Baryon + lepton
number violating processes are in thermal
equilibrium before the electroweak
phase transition, so a primordial asymmetry
must be  stored in a conserved  quantum
number (which could be one of the $B/3 - L_i)$ to
survive. For  $B/3 - L_e$ to be
effectively conserved before the
phase transition, we need \cite{cdeo}
$(B_4{\mu}^e - \mu_4 {B}^e) \lappeq 2 \times 10^{-7} |\mu| |B|/\tan \beta $.
\footnote{We have added a factor of $\tan \beta$ with respect to
the first reference of \cite{cdeo}. This comes from requiring that
the lepton number violation due to the soft masses
be out of equilibrium; the
minimisation conditions of the scalar potential say that
the RPV soft masses $\tilde{m^2}_{4j} = -  B_j \tan \beta$,
so can be larger than the $B$ terms for large $\tan \beta$.}
For $\delta_{\mu} \sim 10^{-6}$, 
 $\delta_{B} \sim 10^{-4}$, as could
generate the atmospheric and 
SMA solar masses, this
is satisfied for 
$\hat{\delta}_{\mu}^e \lappeq .1$ or
$\hat{\delta}_{B}^e \lappeq 2 \times 10^{-3}$.

In the case of a hierarchical spectrum, then only for very small 
values of the angle $\gamma$  
do we find that $\cos^2\gamma m_2 \sim \sin^2\gamma m_3$.
In this case $|\delta_{\mu}| $ will generate the neutrino with 
a mass $m_3 \rightarrow \sqrt {\Delta m_{atm}^2}$ and  $|\delta_{B}| $ 
will generate the neutrino with a mass $m_2 \rightarrow
 \sqrt{\Delta m_{solar}^2}$. For all other values of $\gamma$ then 
 $|\delta_{B}| \rightarrow \sqrt {\Delta m_{atm}^2} $ and  $|\delta_{\mu}| 
 \rightarrow \sqrt{\Delta m_{solar}^2 }$. That is,
the loop is giving the largest contribution to the neutrino masses. 
So the ``measure'' on parameter space determined by the angle $\gamma$
does not correspond to natural theoretical expectations.
For large values of $\gamma$  the atmospheric mixing
angle is now controlled by the relative size of 
$\delta_{B_{\perp}}^{\mu}$ and $\delta_{B_{\perp}}^{\tau}$.

\subsection{ 
Masses and mixing angles from
$\vec{\delta}_\mu$ and $\vec{\delta}_B$ }

\begin{figure}[hbt]
\begin{center}
\vspace{-30pt}
\hspace{0.2cm}\epsfxsize=16cm\epsfbox{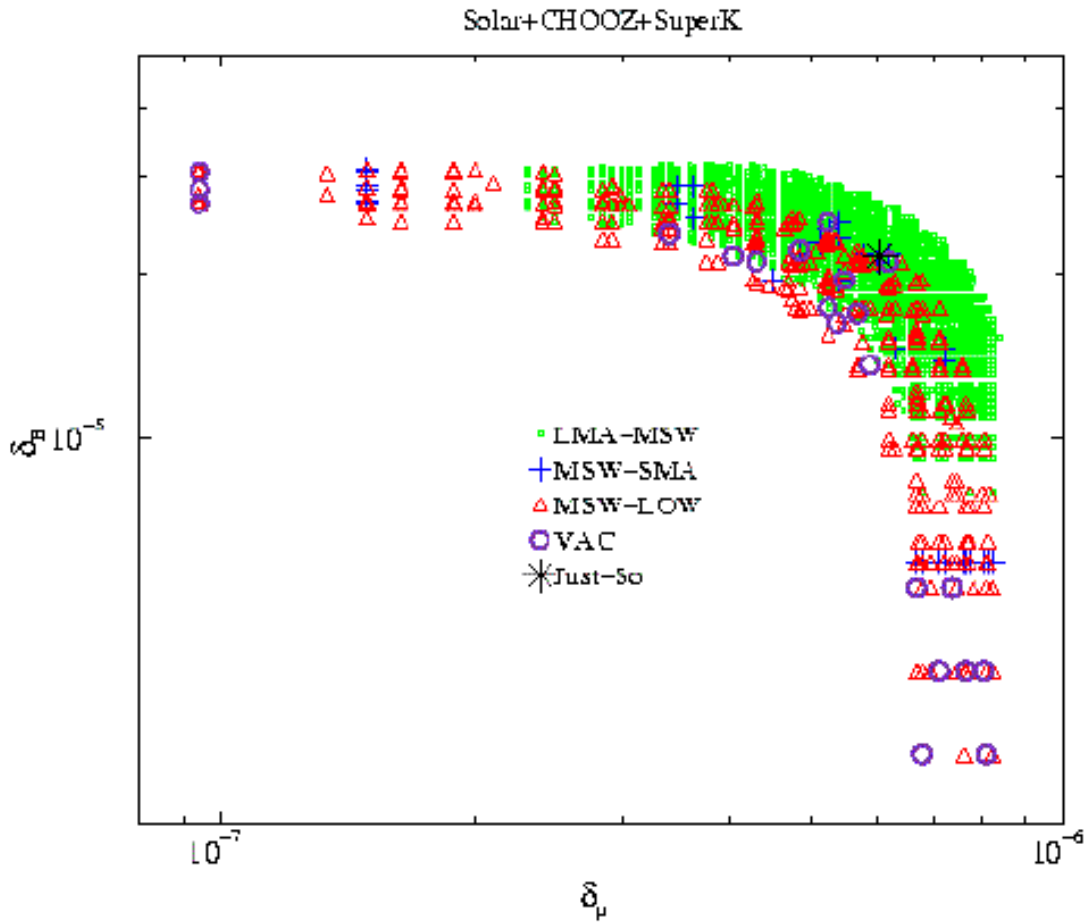}

\caption{We present the available region in parameter space for a combined 
fit using Chooz, atmospheric and one of the solar
oscillations solutions.}
\protect\label{dmuvsdB}
\end{center}
\end{figure}

In this subsection we present numerical results of our 
scan of parameter space 
varying the inputs parameters in the range
of $|\delta_{\mu}| <10^{-6}$ and $|\delta{_B}| < 
4.10^{-5}$. This scan allows us to 
consider the case when tree contributions are larger, 
of the same order or smaller
than the loop contributions. We present in figure 
\ref{dmuvsdB} the allowed region 
in parameter space in
the  $|\delta_{B}|$ vs  $|\delta_{\mu}|$ plane,
 these results are in agreement with the ones obtained in ref. \cite{DL2}. 
 The different points displayed correspond to  combined constraints from 
 Chooz+SuperK atmospheric data and one of the solar solutions. As expected,
figures \ref{dmuvsdB} and \ref{fLMA} are quite similar. \\
 
\begin{figure}[h!]
\vspace{-40pt}\begin{center}
\centerline{\hspace{-3.3mm}}
\hspace{-0.1cm}
\epsfxsize=14cm\epsfbox{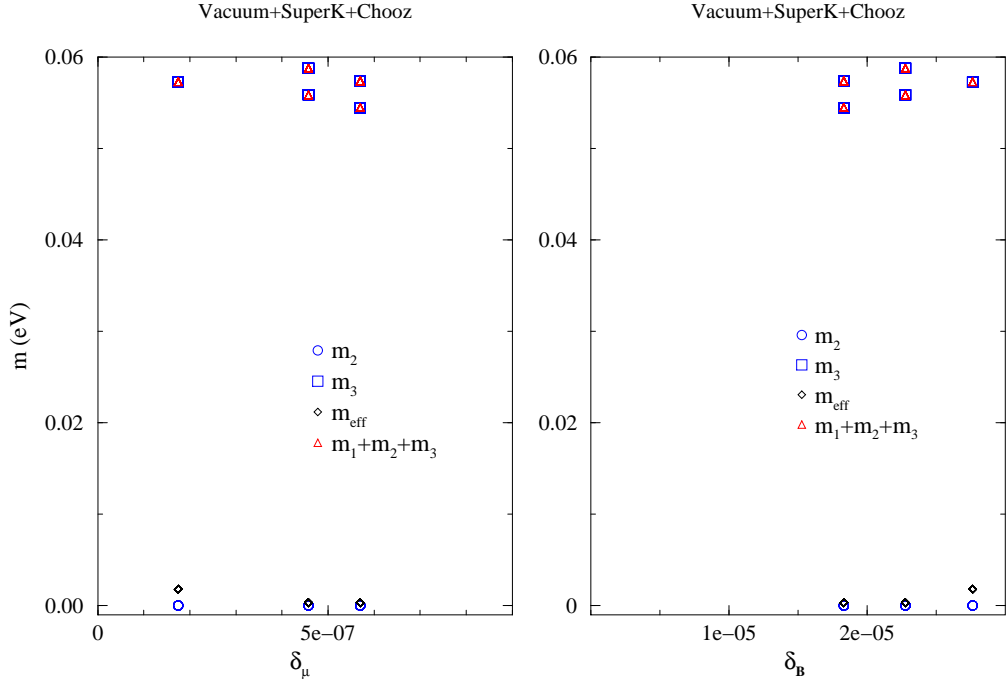}
\caption{The mass spectrum is represented together with $m_{\mathrm{eff}}$ and
the sum of the eigenvalues in the case of a combined fit with Chooz, Vacuum
solar solution and atmospheric SuperK constraints.}
\protect\label{spec-vac}\end{center}
\end{figure}

\begin{figure}[h!]
\vspace{-40pt}
\begin{center}
\centerline{\hspace{-3.3mm}}
\hspace{-0.1cm}
\epsfxsize=14cm\epsfbox{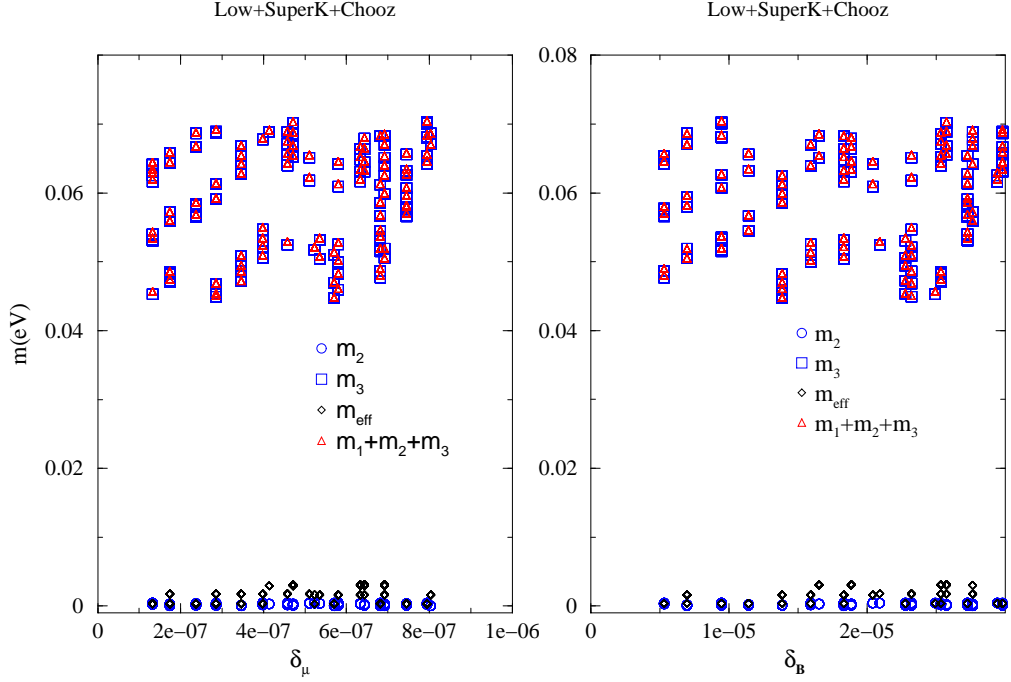}
\caption{The mass spectrum together with the effective mass $m_{\mathrm{eff}}$
and the sum of the eigenvalues are represented for a combined fit with 
Chooz, Superk and MSW-LOW oscillation 
solutions. }
\protect\label{spec-low}\end{center}
\end{figure}

\begin{figure}[h!]
\vspace{-40pt}\begin{center}
\centerline{\hspace{-3.3mm}}
\hspace{-0.1cm}
\epsfxsize=14cm\epsfbox{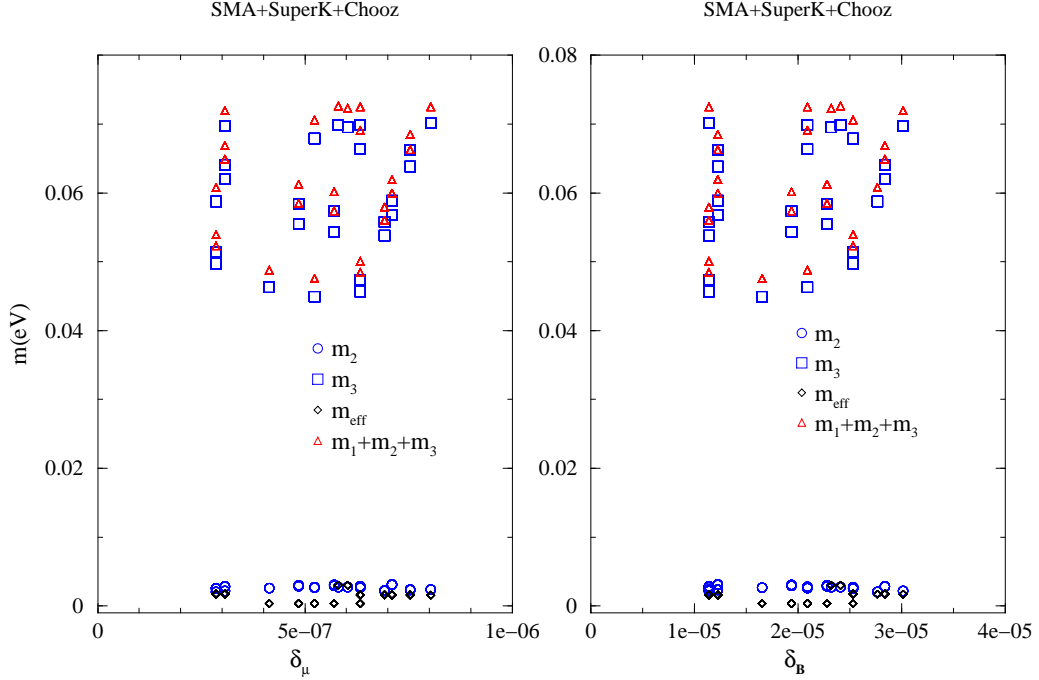}
\caption{The mass spectrum together with the effective mass $m_{\mathrm{eff}}$
and the sum of the eigenvalues are represented in the case of  a combined 
fit with 
Chooz, Superk atmospheric and MSW-SMA oscillation 
solutions. }
\protect\label{spec-sma}\end{center}
\end{figure}

\begin{figure}[hbt]
\vspace{-40pt}\begin{center}
\centerline{\hspace{-3.3mm}}
\hspace{-0.1cm}
\epsfysize=9cm\epsfbox{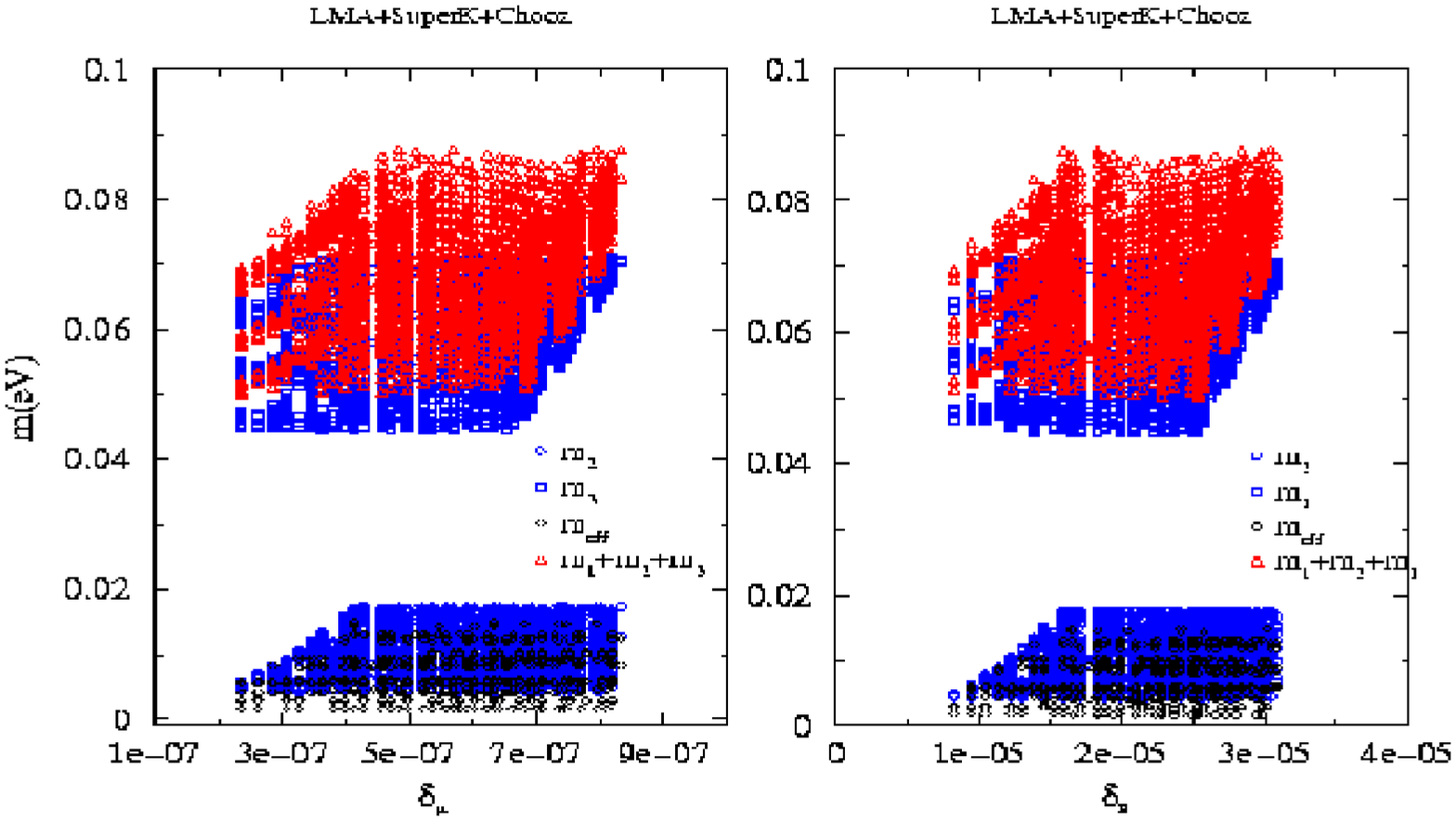}
\caption{The mass spectrum together with the effective mass $m_{\mathrm{eff}}$
and the sum of the eigenvalues are represented for a combined fit with 
Chooz, Superk atmospheric and MSW-LMA constraints. }
\protect\label{spec-lma}\end{center}
\end{figure}

In order to study the spectrum for each of the possible combined solutions we
give samples of the different mass spectra for different values of the input
parameters $\vec\delta_{\mu}$ and $\vec\delta_B$. For illustration, 
in figures \ref{spec-vac},\ref{spec-low}, \ref{spec-sma} and \ref{spec-lma} 
 we present the obtained spectra for the combined 
constraint of 
Chooz and the atmospheric SuperK data with one of the solar solutions 
as functions 
of  $|\delta_{\mu}|$ and $|\delta_B|$
. We also plot
$m_{eff}$ and $\sum_i m_{i}$ and we see that the bounds from 
neutrinoless double beta decay \cite{76Ge} and the cosmological 
constraint $\sum_i m_{i}\lsim
$few eV are  respectively fulfilled.
From these plots one can read that the scan 
shows us only hierarchical spectra 
 for all combined constraints.

\begin{figure}[h!]\begin{center}
\centerline{\hspace{-3.3mm}}
\hspace{-0.1cm}\epsfxsize=12cm\epsfbox{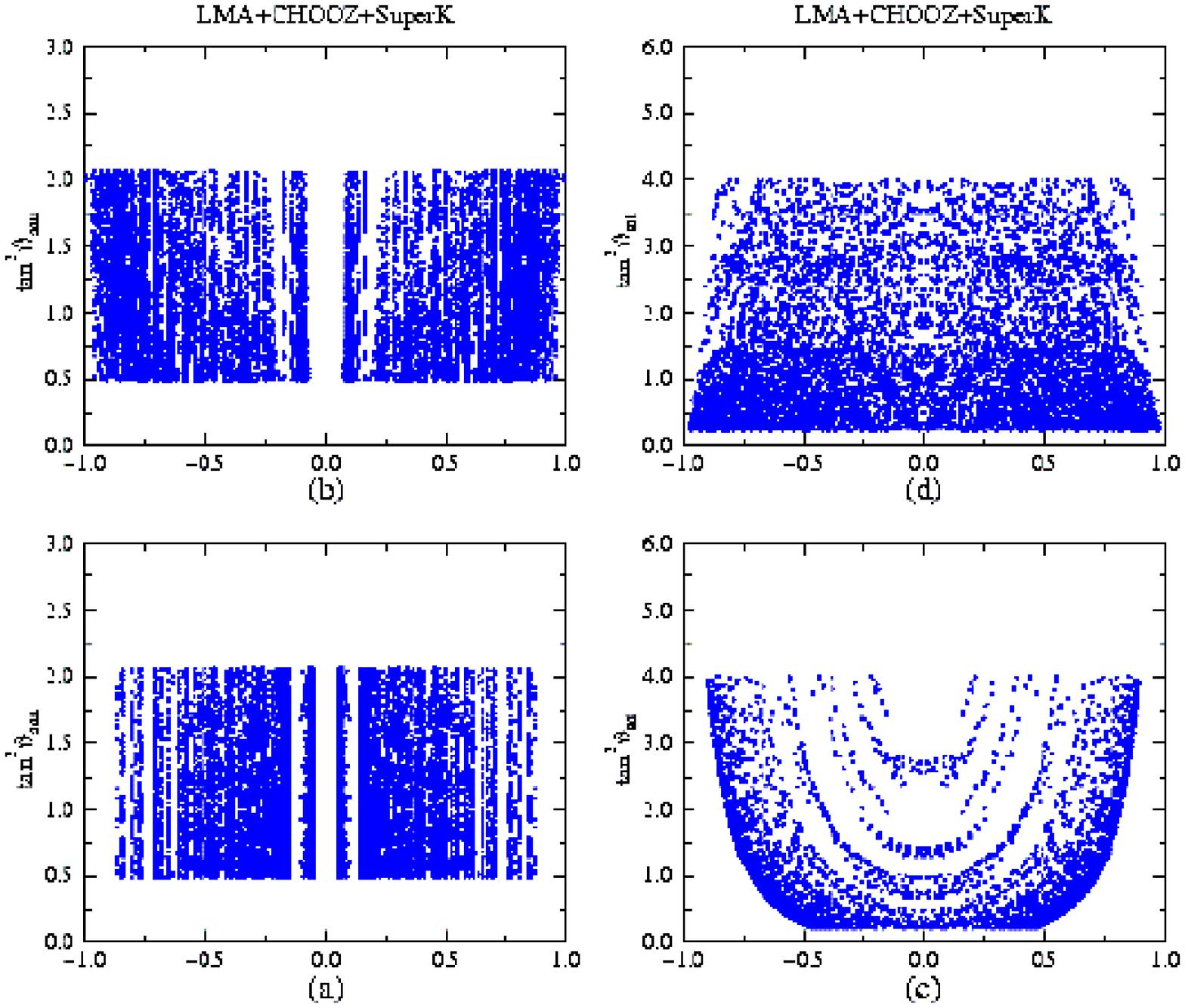}
\hspace{0.3cm}
\caption{The allowed atmospheric and solar angles for a combined fit of 
MSW-LMA, Chooz and SuperK atmospheric data. The atmospheric (solar) angle is
 plotted as a
function of ${|\hat{\delta}_{\mu}^{1}|}$  and  ${|\hat{\delta}_{\mu}^{2}|}$ (${|\hat{\delta}_{B_{\perp}}^{1}|}$
 and ${|\hat{\delta}_{B_{\perp}}^{2}|}$) in (a) and (b) (in
 (c) and (d)) respectively.}
\protect\label{angleslma}\end{center}
\end{figure}

\begin{figure}[h!]\begin{center}
\centerline{\hspace{-3.3mm}}
\hspace{-0.1cm}\epsfxsize=12cm\epsfbox{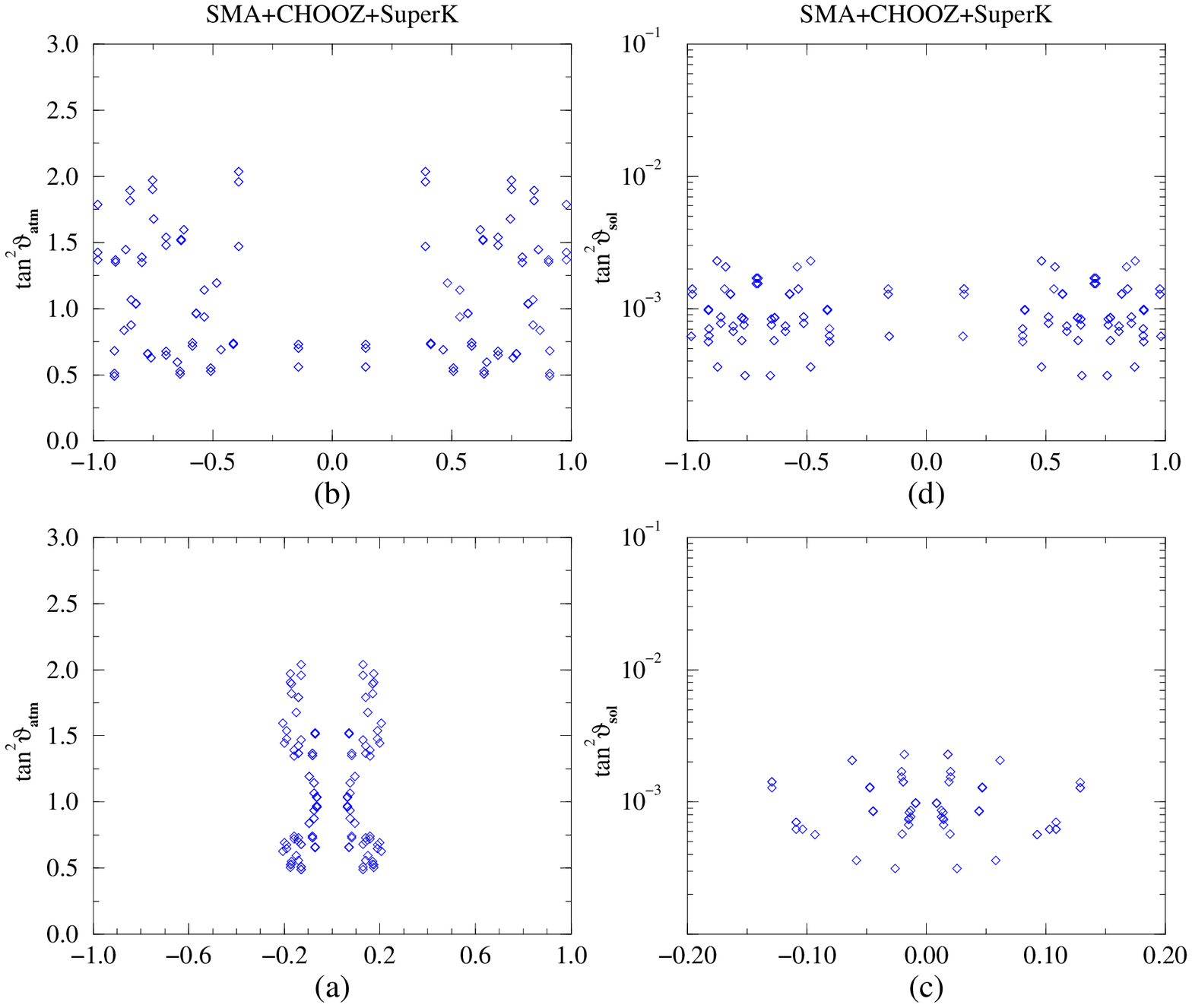}
\hspace{0.3cm}
\caption{The allowed atmospheric and solar angles for a combined fit of 
MSW-SMA, Chooz and SuperK atmospheric data. The atmospheric (solar) angle is
 plotted as a
function of ${|\hat{\delta}_{\mu}^{1}|}$  and  ${|\hat{\delta}_{\mu}^{2}|}$ (${|\hat{\delta}_{B_{\perp}}^{1}|}$
 and ${|\hat{\delta}_{B_{\perp}}^{2}|}$) in (a) and (b) (in
 (c) and (d)) respectively. }
\protect\label{anglessma}\end{center}
\end{figure}

\begin{figure}[h!]\begin{center}
\centerline{\hspace{-3.3mm}}
\hspace{-0.1cm}\epsfxsize=12cm\epsfbox{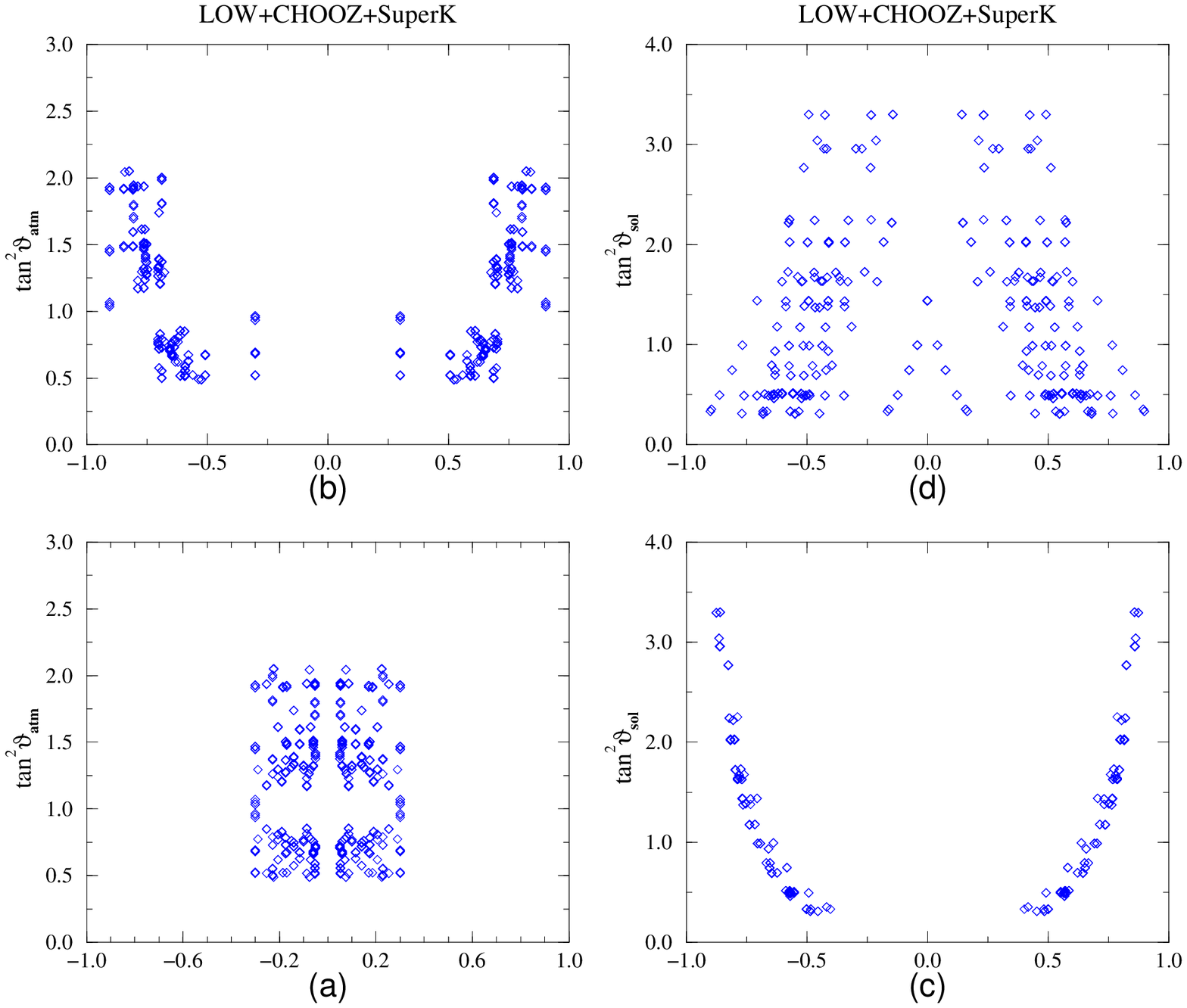}
\hspace{0.3cm}
\caption{The allowed atmospheric and solar angles for a combined fit 
of MSW-LOW, Chooz and SuperK atmospheric constraints. The atmospheric (solar) angle is plotted as a
function of ${|\hat{\delta}_{\mu}^{1}|}$  and  ${|\hat{\delta}_{\mu}^{2}|}$ (${|\hat{\delta}_{B_{\perp}}^{1}|}$
 and ${|\hat{\delta}_{B_{\perp}}^{2}|}$) in (a) and (b) (in
 (c) and (d)) respectively. }
\protect\label{angleslow}\end{center}
\end{figure}

\begin{figure}[h!]\begin{center}
\centerline{\hspace{-3.3mm}}
\hspace{-0.1cm}\centerline{\hspace{-3.3mm}}
\hspace{-0.1cm}\epsfxsize=12cm\epsfbox{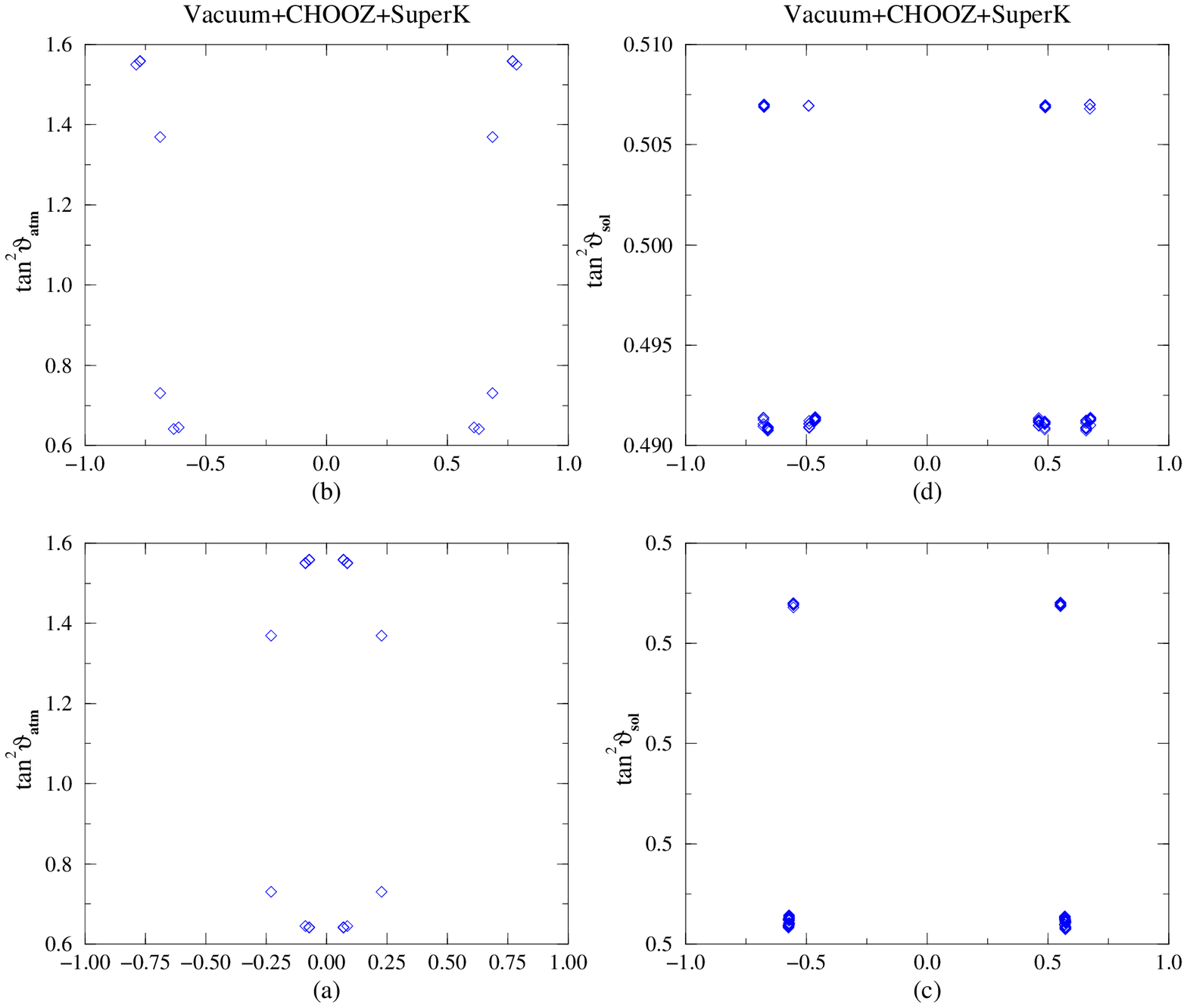}
\vspace{2cm}
\caption{The allowed atmospheric and solar angles for a combined fit of 
Vacuum, Chooz and SuperK atmospheric solutions.The atmospheric (solar) angle is plotted as a
function of ${|\hat{\delta}_{\mu}^{1}|}$  and  ${|\hat{\delta}_{\mu}^{2}|}$ (${|\hat{\delta}_{B_{\perp}}^{1}|}$
 and ${|\hat{\delta}_{B_{\perp}}^{2}|}$) in (a) and (b) (in
 (c) and (d)) respectively.  }
\protect\label{anglesvac}\end{center}
\end{figure}

In figures \ref{angleslma}, \ref{anglessma}, \ref{angleslow} and \ref{anglesvac}
 we plot the different solar and atmospheric mixing angles as functions of the 
 R-parity violating parameters  ${|\delta_{\mu}^{1,2}|\over |\delta_{\mu}|}$
and  ${|\delta_{B_{\perp}}^{1,2}|\over |\delta_{B_{\perp}}|}$. The
allowed ranges for the RPV parameters agree with the estimates
of equations (\ref{delm}) and (\ref{delBperp}).

In general, the observed discreteness in all our plots is in part due 
to the step taken
 in our scan of parameter space and we note that
  the dependence on a single parameter is complicated since many 
  solutions can be found for the same input value. Also the same value
of the physical parameter involved can correspond to different values
of the inputs. 

\section{Model with only $\delta_{\mu}$}
\label{sec3}

In this section, we show that a realistic
neutrino mass matrix cannot be generated in an
RPV model where 
only the invariant $\vec{\delta}_{\mu}$ is non-zero.
This toy model is interesting because it allows us to study 
whether the $O( |\delta_{\mu}|^2)$
 loop contributions to $ \mnu$ need to be taken into account.
We assume that the only non-zero invariant is  $\vec{\delta}_{\mu}$,
and ask whether we can generate $m_2$ from the 
$\delta_{\mu}$ loops of figure  \ref{fglambda} if $m_3$
is due to the tree level diagram of figure \ref{tree}.
 The answer is no, even for large $\tan \beta$, because
the loop mass is necessarily small (of order
the vacuum or just-so masses), and the solar mixing
angle is also small. 

The analysis of this section is quite relevant as this type of
model has been  constrained using other low energy LFV processes, as in
\cite{FH}. However, as mentioned in the introduction, typically
neutrino data will put tighter constraints on the RPV parameters and
as we have seen it is not possible to accomodate all neutrino data
simultaneously.

We take 
\beq
|\vec{\delta}_{\mu}| \sim \frac{[\Delta m^2_{atm}]^{1/4}}{m_{SUSY}}
\eeq
and $\delta_{\mu}^{\mu}\sim \delta_{\mu}^{\tau} \gg
\delta_{\mu}^{e}$  so that the heaviest neutrino
mass $m_3$ explains the atmospheric neutrino
anomaly and is consistent with Chooz.  The
solar neutrino mass will come from the ``see-saw-type'' loops
of figure \ref{fglambda}, with
one $\delta_{\mu}$ mass insertion on an external leg.

 The  diagrams of figure \ref{fglambda}
can have R-parity violating mass insertions at
(I,V), (I,III) or (III,VIII).
The potentially largest is (I,V)
 (see \cite{DL2} for exact formulae),
because it is proportional to $\tan \beta$:
\beq
 (I,V) \sim g \delta_{\mu}^i  \delta_{\mu}^j  \tan \beta 
\frac{m_{e_j}^2}{16 \pi^2 m_{SUSY}} ~~~~.
\label{IV}
\eeq
We refer to this as ``see-saw-type'' because
it is a mass between the neutrino $\nu_3$ who
acquired mass at tree level, and the 
massless-at-tree-level neutrinos $\nu_2$ and $\nu_1$. Neglecting
the muon and electron masses with respect
to the tau mass, equation (\ref{IV}) is a mass
between $\nu_3$ and $\nu_\tau$.
The largest  see-saw mass it could induce for  $\nu_2$ 
would  arise if $\nu_2 \sim \nu_{\tau}$, and
is of order
\beq
m_2 \lsim m_\nu^{tree} \left( \frac{\tan \beta
h_\tau^2}{8 \pi^2} \right)^2 ~~~.
\eeq
This could generate $m_2 \sim 10^{-5} m_\nu^{tree}$ 
for $\tan \beta \sim 50$.  So if all free parameters
are stretched, it might just be possible to
get a ``see-saw'' loop mass of order the Vacuum
or JUST SO solar neutrino masses.

The neutrino which gets a loop mass will
only have a very small $\nu_e$ component, so
will not have the large mixing angle required
for the Vacuum and JUST SO solar solutions.
However, recall that we  choose $\delta_{\mu}^{\tau} \sim \delta_{\mu}^{\mu} \gg
\delta_{\mu}^{e}$, so that the tree level mass
explains the atmospheric neutrino anomaly
and agrees with Chooz. 
The loops mix $\nu_3$ with $\nu_\tau$, so
the second massive neutrino will be composed
largely of $\nu_\tau$ and $\nu_\mu$, and the
massless neutrino will be mostly $\nu_e$. An
$R_p$ violating model with $\delta_\mu \neq 0$ and
all the other invariants zero  therefore
can not explain observed neutrino data.

\section{Effects on low energy LFV processes}

It is also of interest  to comment on
the effects of the allowed values for
the different $\delta_{B}^{i}$, $\delta_{\mu}^{i}$  from
 neutrino masses and mixings on 
low energy lepton flavour violating processes. In this model
there  can be tree level contributions from the RPV parameters
to some of the $\Delta L =0$ processes.
We can estimate the contributions
to processes like $\mu \rightarrow e \gamma$, $\mu\rightarrow eee$
$\mu - e$ conversion in nuclei, and other LFV processes at
low energy.

In general the branching ratios for these processes will be proportional to

 \beq
 \left({\delta_{\mu}^{i}\delta_{\mu}^{j}\over G_f m_{susy}^2}\right)^2, \hspace{1cm}
 \left({\delta_{B}^{i}\delta_{B}^{j}\over G_f m_{susy}^2}\right)^2, \hspace{1cm}
 \left({\delta_{\mu}^{i}\delta_{B}^{j}\over G_f m_{susy}^2}\right)^2.
 \eeq

Thus it is trivial to see using simple estimates
that the branching ratios 
of low energy lepton flavour violating
processes will  automatically be suppressed given the constraints from
neutrino data.
 The obtained branching ratios are
 well below current  and anticipated experimental values.

Our simple model for generating neutrino masses and mixings
correlates the ratios of  branching ratios of different LFV low
energy processes. The exact values of these correlations will
depend on the combined constraint of solar, atmospheric and
Chooz experiments which is used.

A generic consequence of this model is that the lightest supersymmetric
particle (LSP)
will decay. The details of this analysis will be presented elsewhere.

\section{SUMMARY}

In this paper, we studied a minimal model of bilinear
$R_p$ violation which is consistent with the atmospheric and
solar neutrino anomalies. In the ``basis-independent'' formalism
in which we work, the model has $\vec{\delta}_\mu \neq
\vec{\delta}_B \neq 0$, and all other invariants zero 
\footnote{The vectors and lower case indices are in lepton flavour
 space; our notation is explained in the introduction and in table
 \ref{tinvar}.}.
 In the basis where there are no $\Rpv$ trilinears,
this  model has R-parity
violating   $B_i H_uL^i$ masses
and sneutrino vevs. Such a model could arise
if the superpotential was $R_p$ conserving, but
R-parity violation was introduced through the soft terms.

In the limit where we neglect MSSM mixing angles and
set all spartner masses to a common scale $m_{SUSY}$,
the neutrino mass matrix is controlled by the
6 parameters $\delta^i_\mu, \delta^i_B$ (i: 1..3).
 There are two massive neutrinos.
If we make the simplifying assumption that we
know the neutrino masses and the mixing matrix (MNS) exactly,
we can solve for  $\vec{\delta}_\mu$ and $\vec{\delta}_B$
as a function of  one additional parameter,
which can be taken as the angle between 
$\vec{\delta}_\mu$ and $\vec{\delta}_B$.
We plot the allowed values of  $|\vec{\delta}_\mu|$
versus $|\vec{ \delta}_B|$ for different solar solutions
in figure \ref{fLMA}. See also figure \ref{dmuvsdB}. This gives
some idea of the weak-scale $\Rpv$ Lagrangian mass terms 
which can induce a phenomenologically correct neutrino sector.
 In reality
however, we do not know the  neutrino masses or
the MNS matrix elements exactly, so we numerically vary the six inputs parameters,
and find ranges where the  neutrino masses and MNS matrix elements 
 are consistent with combined fits of
SuperK, Chooz and solar data.

This model differs from usual models of 
bilinear $\Rpv$ in that it  has different loop
diagrams.  $\vec{\delta}_B$
generates a loop contribution to
the neutrino mass matrix of order $g^2 |\vec{\delta}_B|^2
m_{SUSY}$ via the Grossman-Haber diagram of
figure \ref{fGH}. This diagram has frequently
been overlooked in the literature. 
If $\vec{\delta}_\mu \propto \vec{\delta}_B$,
then this is a loop correction to
the tree mass and so its not useful to generate another
neutrino mass. 
However  we treat  $\vec{\delta}_\mu$  
and $\vec{\delta}_B$ as phenomenological
weak-scale inputs,  and allow them to vary
independently. The Grossman-Haber diagram
 is therefore the  source of
our loop neutrino masses.
 This is in contrast to a MSUGRA bilinear model,  
where $\mu$ is misaligned at the GUT scale with 
respect to the trilinears, and the
 loop diagrams are proportional to  trilinear
couplings squared. The trilinear diagrams are more important
than the Grossman-Haber diagram in MSUGRA models, 
 because renormalisation
group running  makes the latter
  a loop correction to the tree level mass 
($\vec{\delta}_\mu \sim \vec{\delta}_B$).

The simplicity of this $\vec{\delta}_\mu -
 \vec{\delta}_B$ model is attractive:
neutrino oscillations can be explained with
six additional parameters (and all of the MSSM!).
 This is in contrast to most models with trilinear
R-parity violation, where a multitude
of couplings contribute to the neutrino masses.
The phenomenology of this model is quite different
from other RPV violating models with respect to the
scalar sector of the theory. 
For neutrino masses
$m_3 = \sqrt{\Delta m^2_{atm}}$ and 
$m_2 = \sqrt{\Delta m^2_{sol}}$ and  a given MNS matrix,  the
$\Rpv$  mass terms $B_i$ and $\mu_i$
in the Lagrangian are determined
as a function of one free parameter,
which is the angle $\gamma$ in equations
(\ref{mb0}) to (\ref{delBperp}). This
parametrisation can describe more general
models of bilinear $\Rpv$ under certain circumstances,
as was discussed in section \ref{dmudB}.
We have also shown that combined solutions of SuperK, 
Chooz and all possible solar oscillations solutions can be obtained.

\section*{Acknowledgements}

We would like to thank H. Haber for discussions.
The work of M.L. was partially supported
by Colciencias-BID, under contracts nos. 120-2000 and 348-2000.
M. L. would like to thank ICTP-Trieste for hospitality during the completion
of this work.

\end{document}